\documentclass[twoside,twocolumn,9pt]{article}
\usepackage{extsizes}
\usepackage[super,sort&compress,comma]{natbib} 
\usepackage[version=3]{mhchem}
\usepackage[left=1.5cm, right=1.5cm, top=1.785cm, bottom=2.0cm]{geometry}
\usepackage{balance}
\usepackage{times,mathptmx}
\usepackage{sectsty}
\usepackage{graphicx} 
\usepackage{lastpage}
\usepackage[format=plain,justification=justified,singlelinecheck=false,font={stretch=1.125,small,sf},labelfont=bf,labelsep=space]{caption}
\usepackage{float}
\usepackage{fancyhdr}
\usepackage{fnpos}
\usepackage[english]{babel}
\addto{\captionsenglish}{%
  
}
\usepackage{array}
\usepackage{droidsans}
\usepackage{charter}
\usepackage[T1]{fontenc}
\usepackage[usenames,dvipsnames]{xcolor}
\usepackage{setspace}
\usepackage[compact]{titlesec}
\usepackage{hyperref}

\usepackage{epstopdf}

\definecolor{cream}{RGB}{222,217,201}

\begin{document}

\pagestyle{fancy}
\thispagestyle{plain}
\fancypagestyle{plain}{

\renewcommand{\headrulewidth}{0pt}
}

\makeFNbottom
\makeatletter
\renewcommand\LARGE{\@setfontsize\LARGE{15pt}{17}}
\renewcommand\Large{\@setfontsize\Large{12pt}{14}}
\renewcommand\large{\@setfontsize\large{10pt}{12}}
\renewcommand\footnotesize{\@setfontsize\footnotesize{7pt}{10}}
\makeatother

\renewcommand{\thefootnote}{\fnsymbol{footnote}}
\renewcommand\footnoterule{\vspace*{1pt}%
\color{cream}\hrule width 3.5in height 0.4pt \color{black}\vspace*{5pt}} 
\setcounter{secnumdepth}{5}

\makeatletter 
\renewcommand\@biblabel[1]{#1}            
\renewcommand\@makefntext[1]%
{\noindent\makebox[0pt][r]{\@thefnmark\,}#1}
\makeatother 
\renewcommand{\figurename}{\small{Fig.}~}
\sectionfont{\sffamily\Large}
\subsectionfont{\normalsize}
\subsubsectionfont{\bf}
\setstretch{1.125} 
\setlength{\skip\footins}{0.8cm}
\setlength{\footnotesep}{0.25cm}
\setlength{\jot}{10pt}
\titlespacing*{\section}{0pt}{4pt}{4pt}
\titlespacing*{\subsection}{0pt}{15pt}{1pt}

\fancyfoot{}
\fancyfoot[RO]{\footnotesize{\sffamily{1--\pageref{LastPage} ~\textbar  \hspace{2pt}\thepage}}}
\fancyfoot[LE]{\footnotesize{\sffamily{\thepage~\textbar\hspace{3.45cm} 1--\pageref{LastPage}}}}
\fancyhead{}
\renewcommand{\headrulewidth}{0pt} 
\renewcommand{\footrulewidth}{0pt}
\setlength{\arrayrulewidth}{1pt}
\setlength{\columnsep}{6.5mm}
\setlength\bibsep{1pt}

\newcommand{\JH}[1]{\textcolor{blue}{JH: #1}}
\newcommand{\HM}[1]{\textcolor{green}{HM: #1}}

\makeatletter 
\newlength{\figrulesep} 
\setlength{\figrulesep}{0.5\textfloatsep} 

\newcommand{\topfigrule}{\vspace*{-1pt}%
\noindent{\color{cream}\rule[-\figrulesep]{\columnwidth}{1.5pt}} }

\newcommand{\botfigrule}{\vspace*{-2pt}%
\noindent{\color{cream}\rule[\figrulesep]{\columnwidth}{1.5pt}} }

\newcommand{\dblfigrule}{\vspace*{-1pt}%
\noindent{\color{cream}\rule[-\figrulesep]{\textwidth}{1.5pt}} }

\makeatother

\twocolumn[
  \begin{@twocolumnfalse}
\vspace{3cm}
\sffamily
\begin{tabular}{m{4.5cm} p{13.5cm} }

	& \noindent\LARGE{\textbf{Desorption energy of soft particles from a fluid interface$^\dag$}} \\
\vspace{0.3cm} & \vspace{0.3cm} \\
 & \noindent\large{Hadi Mehrabian,\textit{$^{a,b,e}$} Jacco H. Snoeijer\textit{$^{b,a}$} and Jens Harting $^{\ast}$\textit{$^{c,d,a}$}} \\~\\
%
%
	& \noindent\normalsize{
    The efficiency of soft particles to stabilize emulsions is examined by
	measuring their desorption free energy, i.e., the mechanical work required
	to detach the particle from a fluid interface. Here, we consider rubber-like elastic
	as well as microgel particles, using
	coarse-grained molecular dynamics simulations. The energy of desorption is computed
	for two and three-dimensional configurations by means of the mean
	thermodynamic integration method. It is shown that the softness affects the
	particle-interface binding in two opposing directions as compared to
	rigid particles. On the one hand, a soft particle spreads at the
	interface and thereby removes a larger unfavorable liquid-liquid contact
	area compared to rigid particles. On the other hand, softness provides the particle with an additional degree of
	freedom to get reshaped instead of deforming the interface, resulting in a
	smaller restoring force during the detachment. It is shown that the
	first effect prevails so that a soft spherical particle attaches to the fluid interface more strongly than rigid spheres. Finally, we
	consider microgel particles both in the swollen and in the collapsed
	state. Surprisingly, we find that the latter has a larger binding
	energy. All results are rationalised using thermodynamic arguments and
	thereby offer detailed insights into the desorption energy of soft
	particles from fluid interfaces.}

\end{tabular}
\end{@twocolumnfalse} \vspace{0.6cm}]
\renewcommand*\rmdefault{bch}\normalfont\upshape
\rmfamily
\section*{}
\vspace{-1cm}

\footnotetext{
\textit{$^{a}$ Department of Applied Physics, Eindhoven University of Technology, P.O. Box 513, 5600 MB Eindhoven, The Netherlands}\\
\textit{$^{b}$ Physics of Fluids Group and J. M. Burgers Centre for Fluid Dynamics, University of Twente, P.O. Box 217, 7500 AE Enschede, The Netherlands}\\
\textit{$^{c}$ Helmholtz-Institute Erlangen-N\"urnberg for Renewable Energy, Forschungszentrum J\"ulich, F\"urther Str. 248, 90429 Nuremberg, Germany}\\
\textit{$^{d}$ Department of Chemical and Biological Engineering and Department of Physics, Friedrich-Alexander-Universit\"at Erlangen-N\"urnberg, 
F\"{u}rther Stra{\ss}e 248, 90429 N\"{u}rnberg, Germany}\\
\textit{$^{e}$ Chemical Engineering Department, Massachusetts Institute of Technology, Cambridge, MA 02139, USA}\\
}

\section{Introduction}
\begin{figure*}
\centering
\includegraphics[height=8.0cm]{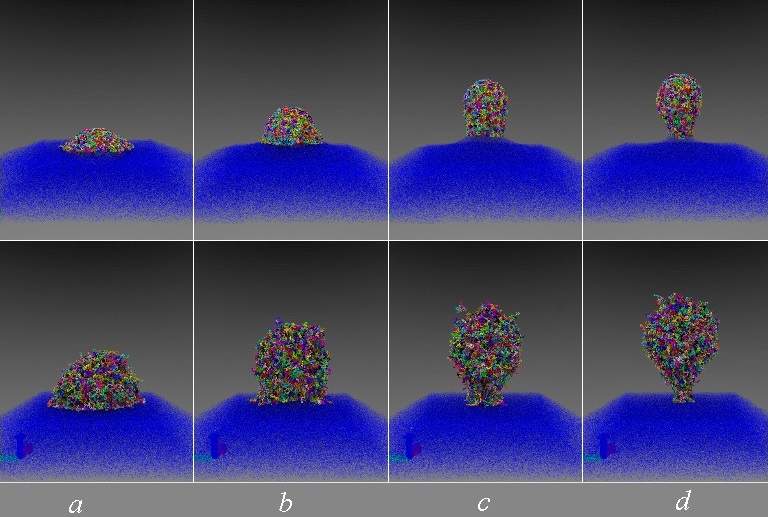}
\caption{The desorption process is illustrated for two types of soft particles,
	as studied here using molecular dynamics simulations. Upper row:
	detachment of an elastic, rubber-like, particle from a flat
	liquid-liquid interface. Lower row: detachment of a swollen microgel
	particle from a flat liquid-liquid interface. Snapshot (a) shows the
	equilibrium state of the adsorbed particle, initially sitting at the
	center of a square fluid interface, and deformed under the act of
	bath surface tension. The particle is quasi-statically moved away from
	the fluid interface until it gets detached (snapshots (b) to (d)). As
	is shown by different colors in the particle, soft particles are made
	from interconnected polymer chains. The blue points depict the
	Lennard-Jones beads for the lower phase, while those for the upper
	phase are not plotted for clarity.}
\label{f:logo}
\end{figure*}
%
%
Emulsions are common in food, pharmaceutical and cosmetics products. In order
to assure their stability and to prevent the separation of the constituting fluid
species, they are often stabilized by adding surfactants. An attractive
alternative to the potentially hazardous surfactants is to add rigid particles
to the emulsion \citep{bib:binks:2002,bib:binks-horozov:2006}, producing
so-called Pickering emulsions \cite{bib:ramsden:1903,bib:pickering:1907}.
Pickering emulsions have few advantages over surfactant stabilized emulsions:
the adsorption of the particles to the interface is nearly irreversible and
they can be made stimulus-responsive by tuning the particle-interface binding
\cite{Jansen2011,Frijters2014}.  Instead of rigid particles, one could think of
soft particles as an alternative for producing Pickering emulsions. Soft
particles, as their name implies, stretch out at the interface and reduce the
unfavorable liquid-liquid contact more efficiently than their rigid
counterparts. It is claimed that the softness of a particle increases the
stability of the emulsion \citep{Destribats2011, Monteillet2014}, while others
observed that softness does not play a major role in stabilizing them
\citep{Brugger2008}.

One aspect of soft particles is their elasticity, allowing them to deform in
their adsorbed state. However, in practice one uses microgels that consist of
an entangled polymeric network into which the solvent molecules can penetrate
\citep{Zhang99,Brugger2009,Saunders1999,Saunders2009}. This process can be
tuned such that a collapse-to-swollen transition takes place when changing e.g.
the temperature or the pH value of the solvent. This makes some of them such as
pNIPAM microgels stimulus responsive, and hence an attractive option for
tunable, reversible stabilization of emulsions. Microgel particles in the
collapsed state behave similar to a generic elastic, rubber-like particle. 
{
In the swollen state, however, their hydrodynamic diameter increases by 2-3 times, their deformability increases \citep{Oh1998}, and their surface properties change.}

It has been observed that the swollen to collapsed transition destabilizes emulsions \citep{Destribats2011},
although the underlying mechanism for this behavior is not fully understood. 


The objective of this paper is to investigate the attachment strength between a
soft particle and a fluid interface, to assess the effectiveness of soft
particles as an emulsifier for Pickering emulsions. This is obviously a key
feature for the stability of Pickering emulsions, but the particle-interface
binding strength has so far not been addressed to the best of our knowledge: a
stronger binding suggests that particles cover the interface more
effectively, and they prefer to remain at the interface under external
fluctuations such as shear or compression forces. The most common way to assess
the binding between a particle and a fluid interface is to measure the energy
required to detach the particle from the interface, the so-called desorption
energy \cite{Rapacchietta1977a,Scheludko1976,Davies2014}. The simplest way to
estimate the binding strength is based on the additional surface energy created
by replacing the particle with the fluid interface. For a rigid particle this
gives \cite{Levine1989b} 
\begin{equation}\label{e:simple}
E=\pi R^2 \gamma_b (1 \pm \cos \theta)^2,
\end{equation}
where $\gamma_b$, $R$, and $\theta$, respectively, are the surface tension of
the fluid interface, the radius of the particle and the particle-interface
contact angle. Using a similar surface area based argument, the desorption
energy of soft particles is calculated by quantifying their shape at the
interface \cite{Style2015}. However, this approach does not account for the
intermediate steps of the desorption process during which the interface and particle get
deformed, producing extra interfacial area -- providing an additional
barrier for the desorption. This could make the desorption energy significantly
larger \cite{Ettelaie2015}. 
In this study, we use computational modeling to characterize the desorption of
soft particles from a liquid-liquid interface. There are few challenges in
modeling the desorption of soft particles which makes most macroscopic computational
modeling techniques inefficient. First, this problem involves contact line
dynamics which is a stress singularity for continuum modeling
methods and different remedies have been developed to address this issue
such as Cahn-Hilliard diffuse interface method (see e.g.
\citep{Mehrabian2011a,Mehrabian2013,Mehrabian2014}). Second, it is nearly
impossible to capture the molecular scale dynamics of microgel particles such as adsorption and solvation of their polymer chains using
any continuum modeling technique. Therefore, a reasonable modelling approach to study the
desorption of soft particles is to use coarse grained molecular dynamics
simulations which can incorporate molecular details of the particle and the
same time capture the macroscopic behaviour. Within the framework of molecular
dynamics simulations, we rely on free energy calculation methods to measure the
desorption free energy. There are two major routes to determine the change in
free energy in molecular mechanics. One approach uses the free energy
perturbation method to create alchemical transformations (see e.g.
\citep{Bellucci,Gumbart2013}), while another one constructs the free energy
profile along a geometrical reaction coordinate using methods such as
thermodynamic integration or umbrella sampling
\citep{WooRoux2005,Mehrabian2018,Mehrabian2019}. 
In this study, we use the thermodynamic integration method to compute the free
energy of binding. We construct the force-distance profile by moving particle
away from the interface in a quasi-static fashion as is illustrated in
Fig.~\ref{f:logo}. The upper and
lower rows, respectively, correspond to collapsed and swollen particles, pulled
from a liquid-liquid interface. Figure~\ref{f:logo}(a) shows the adsorbed state
at equilibrium, while Fig.~\ref{f:logo}(b-d) reveal the intricate deformations
during quasi-static intermediate states of the desorption. From these
simulations, one can reconstruct the force-distance relation and subsequently
the detachment energy is obtained as the mechanical work done by this force
\citep{ChPi2003}, measured by integrating the force acting on the particle
during its removal from the interface. 

In reality, the detachment of a particle from an interface can happen under
different scenarios. For example, the particle can be detached from a flat
interface \citep{Davies2014,Davies2014a} or a curved one
\citep{Ettelaie2015,Hirose2007}, gravity can impose a length scale on the shape
of the interface during the desorption \citep{Obrien1996a}, or the presence of
other particles can affect the way the interface gets deformed \citep{SiHa15}.
In addition, the particle-to-interface size ratio is important especially for
conventional emulsions which contain a range of droplet sizes
\citep{Ettelaie2015}.  
Therefore, it is challenging to provide a
general quantitative theory for the desorption energy, in particular when
particles are deformable. The current study serves as a first step to
investigate the attachment between a soft particle and a fluid interface, and
we therefore restrict ourselves to the simplest possible scenario. Namely, we
ignore the impact of gravity which is a reasonable assumption for emulsions
stabilized by particles with sizes in the range of tens of nanometres to few
micrometres. Additionally, we consider the detachment of a particle from a flat
interface while the liquid always wets the bottom of the container, and we
ignore dynamic (non-equilibrium) effects by moving the particle away from the interface
quasi-statically. 

{
The behavior of microgel particles at the interface, including their swelling/deswelling \citep{Rumyantsev2016}, the effect of compression rate on the surface adsorption \citep{Bochenek2019}, their shape \citep{Camerin2019a} etc. has been studied in the past. However, to the best of our knowledge, none of the previous authors have studied the desorption energy of microgel particles. Style et al. \citep{Style2015} used analytical arguments to measure the desorption energy for an elastic particle, but they admit that such arguments fail for microgel particles, which involves nonlinear dynamics. More recently, molecular dynamics have been used to model nanogels, and the role of interfacial properties and cross-linking density is emphasized \citep{Arismendi-Arrieta2020}.
}

The present study is based on molecular dynamics simulations, as it enables to
capture the internal polymeric structure of microgel particles. However, this
method is computationally costly, and the size of the simulation box should be
chosen judiciously. Most of the results are obtained for cylindrical particles
in a quasi two-dimensional setup, to alleviate the computational cost. Previous
molecular dynamics studies have indeed successfully captured the equilibrium of
wetting and adsorption of deformable particles with liquid and solid surfaces
\citep{Leonforte2011a,Carrillo2010,Cao2015,Cao2014,Mehrabian2016}. For the
desorption, however, the particles need to be pulled from the interface. Here
we measure the desorption energy using the thermodynamic integration method
\citep{Kirkwood1935}, which is commonly used for similar problems
\citep{Razavi2013b,Udayana2010,RaCo14}. 
{The particle is moved away from a flat interface by applying an external force to its center of mass. We define the particle as a single entity in the GROMACS software, and the external force acting on each bead of the particle is proportional to its mass fraction. After each translational step, the position of the center of mass of the particle is restrained by a spring, and the equilibrium force exerted by the interface on the particle is gathered.}

The remainder of the paper is organized as follows. First, we introduce the
problem setup and the key features of the computational method. Then, we derive
the force-distance relationship for the desorption of a rigid cylindrical
particle using the continuum formulation, which serves as a benchmark for our
molecular dynamics simulations. Subsequently, we explore the role of softness
on the desorption of elastic cylindrical particles in their collapsed state.
Next, we study the desorption of elastic spherical particles in a more
realistic three-dimensional configuration, and finally, we discuss the role of
the swollen to collapse transition on the desorption energy of microgel
particles before we conclude the paper. 

\section{Problem setup and governing parameters}\label{s:setup}
In this section, we describe the problem setup and discuss the details of
molecular dynamics simulations and free-energy calculations.
%
%
%
\begin{figure}
 \centering
 \includegraphics[height=4.5cm]{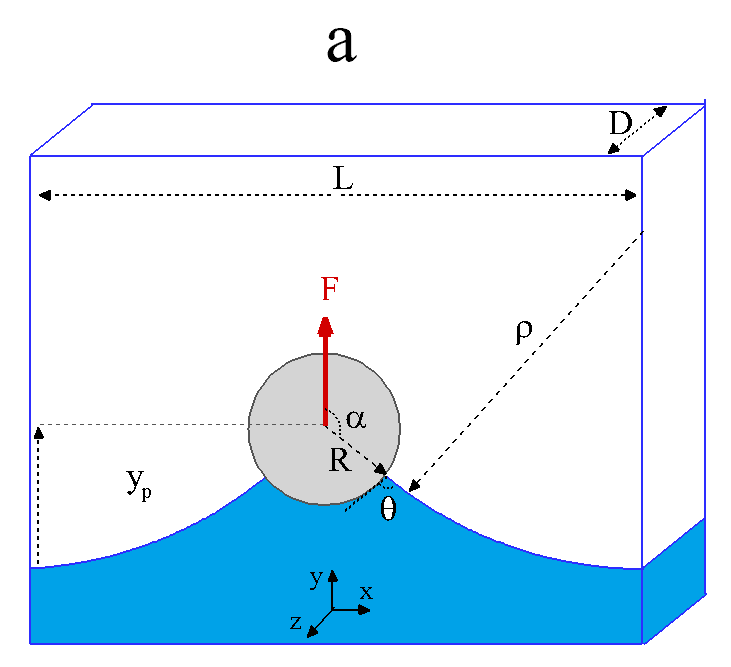}
 \includegraphics[height=3.5cm]{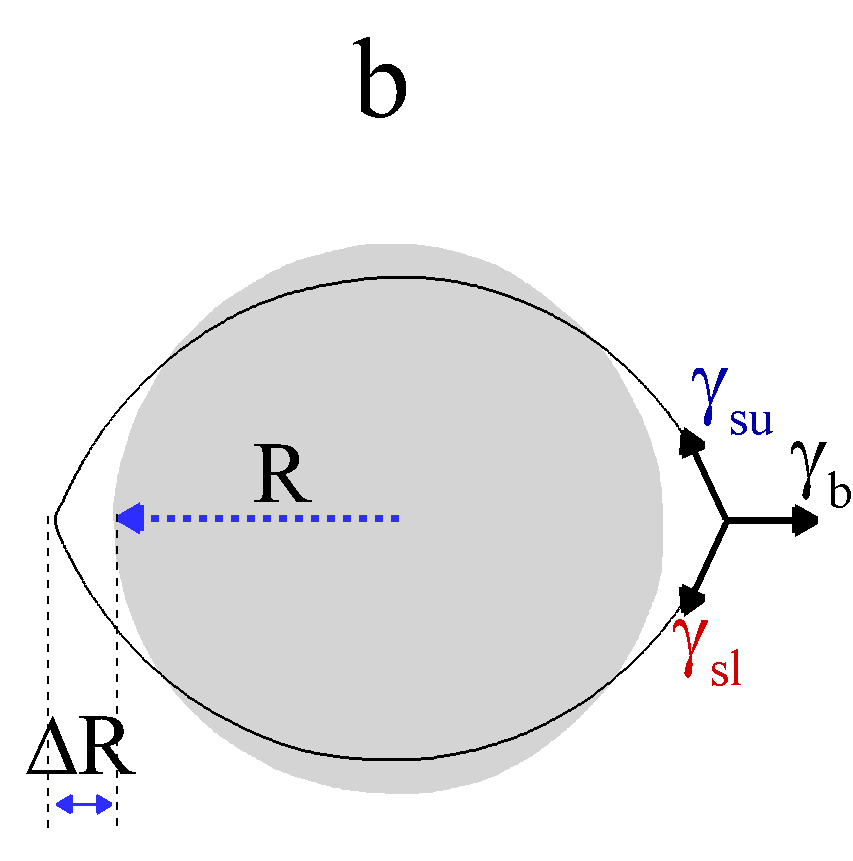}
 \caption{(a) Schematic of the two-dimensional simulation setup. The depth $D$
	and length $L$ of the simulation box are fixed to $0.35R$ (3D
	configuration) and $8.2R$ (2D configuration), respectively. $\rho$ is
	the interface radius of curvature, $y_p$ is the distance of the center
	of mass of the particle from the flat part of the interface, $\alpha$
	is the angle of the particle-interface contact point with the y-axis,
	and $\theta$ is the contact angle. (b) schematic of a soft particle
	deformed at the interface. $R$ is the radius of the undeformed soft
	particle, and $\Delta R$ denotes its deformation.}
 \label{f:setup}
\end{figure}
%
%
%
We study the detachment of soft cylindrical and spherical particles from a
planar fluid interface in two and three-dimensional settings, respectively. The
two-dimensional configuration corresponding to the cylindrical particle is
illustrated in Fig.~\ref{f:setup}(a), in which there are periodic boundary
conditions in the $x$ and $z$ directions. The top and bottom boundaries are
walls to prevent the movement of the system's center of mass in the
y-direction. The two liquid phases have the same density, and the lower and upper liquids
always wets the bottom, and top of the container, respectively.

In this study, we primarily focus on generic incompressible elastic particles without
swelling. For such particles, the problem can be described by seven dimensional
quantities. The radius of the particle, $R$, the length of the liquid bath $L$,
the width of the liquid bath $D$ (Fig.~\ref{f:setup}(a)), Young's elasticity
modulus of the soft particle, $E$, surface tension of the bath, $\gamma_b$, and
the surface tension between the particle and upper and lower liquid phases,
$\gamma_{su}$, and $\gamma_{sl}$, respectively. This will give five
dimensionless groups. At large deformations or when swelling of the particles
becomes relevant, additional information on the molecular level is required
resulting in nonlinear constitutive relations for the macroscopic parameters.

The geometry of the problem is described by two ratios, $L/R$ and $D/R$, which will be kept
fixed during this study. For the two-dimensional configuration, we use
$D/R=0.35$ and $L/R=8.2$. For the three-dimensional configuration
$L/R=D/R=6.4$, while the dimensional radius of the particle is $R=14 \ nm$. The
properties of the particle and of the fluid can be described using three
dimensionless parameters to describe the problem: two softness parameters,
$S=\frac{\gamma_{su}}{ER}$ and $S_b=\frac{\gamma_b}{ER}$, and the wettability
indicated by the contact angle $\theta$. The values of $S$ and $S_b$ can be
seen as the elastocapillary lengths $\gamma_{su}/E$ and $\gamma_b/E$,
normalised by the particle size. The wetting contrast is quantified by the
definition of the Young's angle on rigid surfaces, i.e.
$\cos\theta=\frac{\gamma_{su}-\gamma_{sl}}{\gamma_{b}}$. This makes it possible
to compare rigid and soft particles. All particles in this study are up-down
symmetric except those discussed in Sec.~\ref{s:microgel}. For the symmetric
cases, $\gamma_{su}=\gamma_{sl}$, simply denoted as $\gamma_s$, for which
$\theta=90^\circ$ and $S=\frac{\gamma_s}{ER}$. 

The deformation of the soft particles plays a central role in the current
problem, and has been analysed in detail in our previous work for the case
without external forcing\citep{Mehrabian2016}. It was found that the deformation of a soft cylinder
at an interface with symmetric wetting, quantified by $\Delta R/R$ in
Fig.~\ref{f:setup}(b), is indeed a function of the elastocapillary numbers $S$
and $S_b$. The deformation $\Delta R/R=0$ is zero for rigid particles ($S=0$)
and saturates to a maximum extension as the particle gets softer ($S
\rightarrow \infty$). The exact relationship between the governing parameters
and the deformation is not trivial, but has been characterized in our previous
paper\citep{Mehrabian2016}. 

\section{Molecular dynamics simulations}\label{s:md}

Molecular dynamics simulations are carried out in single precision using the
GROMACS 5 package \citep{VanDerSpoel2005a,Abraham2016}. All restraints were
enforced using the GROMACS pull code, and visualizations are done using the VMD
package \citep{Humphrey1996}.  The two liquid phases consist of beads with
Lennard-Jones interactions. The soft particle is constructed by crosslinking
polymer chains, comprised of bonded beads which interact through Lennard-Jones
interactions with the liquid phases. We use the term ``beads" to refer to the
``coarse-grained particles" in our molecular dynamics simulations in order to
avoid the confusion with the soft particle that is in the focus of this study.
The full details of the molecular dynamics simulations are presented in our
recent paper \citep{Mehrabian2016}. Therefore, here we describe only the main
features of the simulations. 

To keep the model simple, we use the modified shifted and truncated
Lennard-Jones potential \citep{Stecki1995} for the interaction between each
pair of beads
\begin{equation}\label{e:LJ}
U = \left\{
  \begin{array}{ll}
    4\epsilon \left[ \left(\frac{d}{r}\right)^{12}- \left(\frac{d}{r_c}\right)^{12} +\beta\left[\left(\frac{d}{r_c}\right)^6 - \left(\frac{d}{r}\right)^6\right] \right]& \quad \mbox{if $r \leq r_c$}\\
  0 & \quad \mbox{if $r > r_c$,}\\
\end{array} 
\right.
\end{equation}
where $r$ is the distance between two beads, $d$ is the repulsive core diameter
fixed to $0.34$ nm, $\epsilon$ is the depth of the potential minimum set to 3
kJ/mol, and $r_c$ is the the cut-off radius which is equal to $5d$ for the
two-dimensional configuration and $2.5d$ for three-dimensional simulations. We
choose a smaller cut-off radius for the three-dimensional setup in order to
reduce the computational cost. The parameter $\beta$ is used to separate two
liquid phases, and to control the strength of the surface tension between
different phases.  The soft particle is made from the interconnected polymer
chains as mentioned above. Each chain is constructed using a coarse-grained
representation of beads and springs and consists of 32 monomers where two
neighboring beads interact via the following potential:
\begin{equation} \label{eq:FENE}
  U(r)=-\frac{1}{2} k_s R_{max}^2 \ln\left(1-\frac{r^2}{R_{max}^2}\right)+4\epsilon \left(\frac{d}{r}\right)^{12}.
\end{equation}
The first term on the right-hand side is the so-called FENE (finite extensible
nonlinear elastic) potential and controls the amount of the extension. The
second term is the Lennard-Jones repulsion term that accounts for the reduced
volume effect and prevents the collapse of the beads onto each other. We fix
$k_s=25 (\frac{\epsilon}{d^2})$ and $R_{max}= 1.5 d$ to be able to choose a
larger timescale while keeping the FENE links unbreakable \cite{kremer1990}.
The rigidity of the soft particle is controlled by changing the density of the
cross-linking between the chains, and the surface properties are controlled by
changing the Lennard-Jones interactions.

{For all particles in this study, except swollen microgel particles, the bead size is not important as long as it is small compared to the particle length scale. Here, the particle size is around 14 nm, which means that it is around 40 times larger than the bead size. Also, we showed that we could reproduce the macroscopic behavior, which verifies that our choice of length scale is appropriate. Finally, and most importantly, all results are made dimensionless using macroscopic observables such surface tension, elastic modulus, and R in order to highlight the general validity of our results. 
For particles with chain solvation, the solvent/particle, and solvent/solvent bead size could be important. The solvent/particle bead size ratio is smaller than one for most experimental conditions. In this study, we chose the limiting case of the solvent/particle bead size ratio = 1. Also, we used the smallest length scale, which corresponds to the water molecules. Therefore, a particle with a larger bead size would be automatically covered. In addition, in all studied cases, we chose a particle which only wets one phase; we did not try to target a specific two-phase system in which both phases solvate the particle, and hence the solvent/solvent bead size ratio is not important in this study.}

To calculate the force-distance relationship for the desorption of the
particle, we gradually detach the particle from the interface. Particle
transfer from the interface to the liquid bulk is done using constrained
molecular dynamics simulations. The particle's center of mass $y_p$, initially
located at the interface, $y_{p0}$, is the collective variable in this study
and is translated very slowly along the reaction coordinate, which is the
direction normal to the interface, i.e., the y-direction here.  The translation
happens quasi-statically, since the reaction coordinate is divided into smaller
steps whose length is approximately $0.2 \ R$ where $R$ is the radius of the
particle.  After each translational stage, there is an equilibration run of 20
to 30 nanoseconds during which the particle's center of mass is restrained in
space with an umbrella potential, and the average force acting on the particle,
$F$, and the average location $y_p$ are collected. The timestep of MD
simulations is 0.006 picoseconds in these calculations. Then, the
force-distance curve is constructed, and if necessary, more points are added to
refine all areas properly. This process is repeated until the particle gets
detached from the interface and hence, the force exerted on the particle by the
interface becomes zero. The free-energy difference between any two points
during this transfer process can be easily calculated as 
\begin{equation}\label{e:E}
\Delta E = W = -\int_{y_{p0}}^{y_{p}} \langle F \rangle_{y} \ dy,
\end{equation}
where $W$ is the work done on the particle to move it to a distance $y_p$ away
from the interface, $y_{p0}$ is the initial distance of the particle's center
of mass from the interface, and $\langle F \rangle_{y}$ is the ensemble
averaged force in the y-direction acting on the particle at a distance $y_p$
from the interface. Note, that due to the symmetry, the x and z components of
the force are zero. The displacement of the particle, $y_p$, is the distance
between the particle's center of mass and the flat part of the interface at
$x=\pm L/2$, and it is made dimensionless with the particle radius.

\section{Rigid particles \& benchmarking}
%
\begin{figure}
 \centering
 \includegraphics[height=6cm]{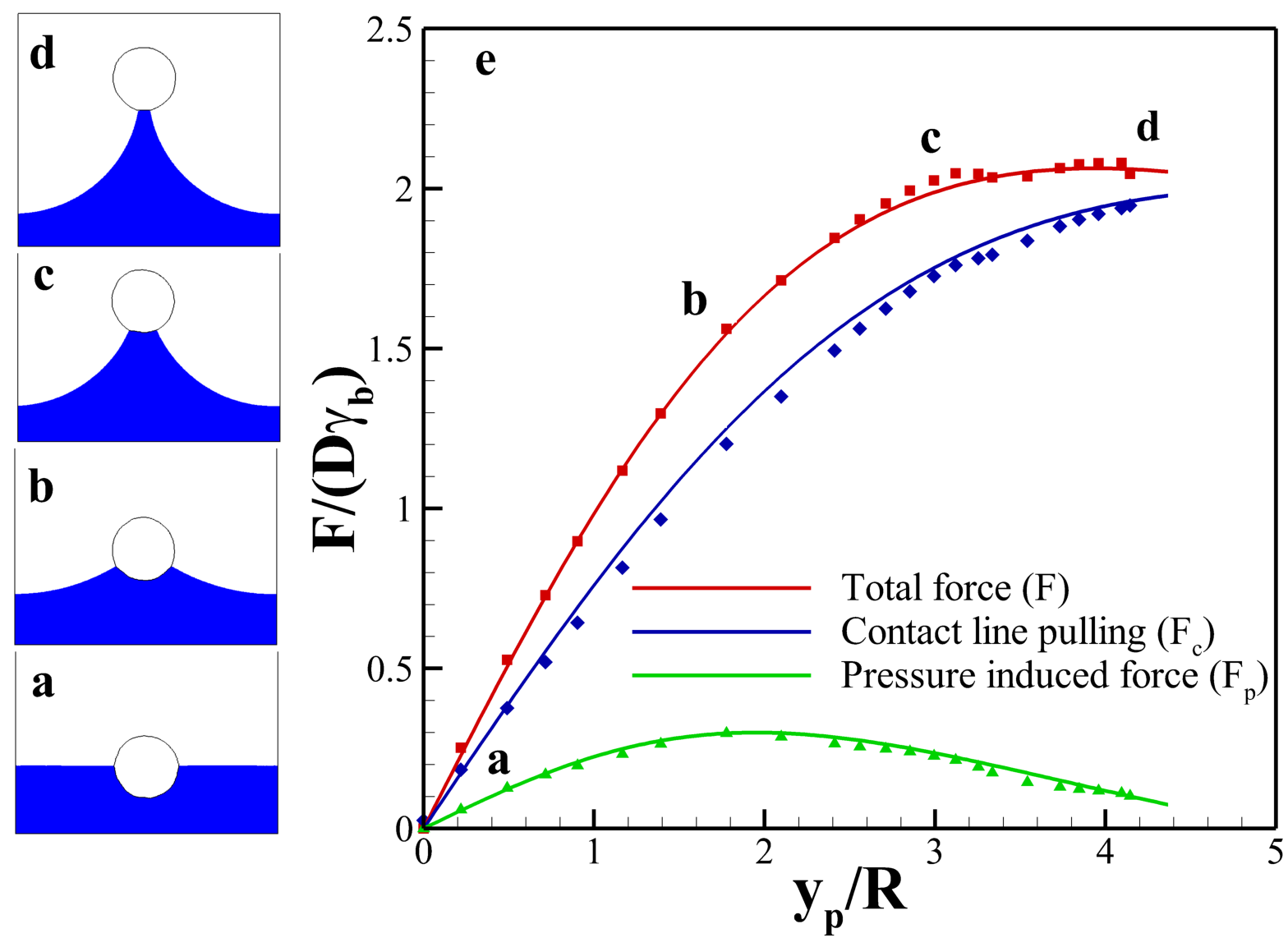}
 \caption{The force-distance relationship for the desorption of a rigid
	cylinder from a fluid interface obtained by molecular dynamics
	simulations and continuum calculation (equations ~\ref{e:force} to
	~\ref{e:yp}) is compared. The total force as well as its components,
	the contact line force ($F_c$) and the pressure force ($F_p$), are
	plotted. The results from molecular dynamics simulations are denoted by
	symbols, while the theoretical curves are given by the solid lines.
	There is an excellent agreement between the two methods which verifies
	the capability of molecular dynamics simulations to capture the
	macroscopic physics of the system.
	}
 \label{f:rigid}
\end{figure}
%
%
%
We consider a rigid cylinder with radius $R$ and  contact angle
$\theta=90^\circ$, initially placed symmetrically at the fluid interface as
is illustrated in Fig.~\ref{f:rigid}(a). To generate a rigid particle in the
molecular dynamics simulations, the density of cross-linking between the
polymer chains is increased to 12 $nm^{-3}$. As is depicted in snapshots (b) to
(d) in Fig.~\ref{f:rigid}, the particle is pulled up quasi-statically until it
gets detached from the fluid interface. Data points in Fig.~\ref{f:rigid}(e) show
the average force and distance calculated using the molecular dynamics
simulations. 

In the absence of gravity and buoyancy forces, the only restoring force acting
on the particle is the force created by the fluid interface, which acts on the
particle in two ways. First, as the particle moves up, the bath interface
becomes more curved whose radius of curvature, $\rho$, produces a negative
capillary pressure below the particle, pulling it down. This component of the
force is negligible for macroscopic particles due to the comparatively small
curvature of the fluid interface. However, for nanoparticles, it has a
considerable effect on the overall force and should be included in the
analysis. We call it the pressure force, $F_p$, as it is obtained by
integrating the pressure around the particle. Second, the liquid interface
pulls the particle down along its contact lines with the particle which we
call the contact line force and denote it by $F_c$, hereafter. The
total restoring force acting on the particle is equal to
\begin{equation}\label{e:force}
F= F_c + F_p = -2 \gamma_b D ( \sin (\alpha- \theta) + \frac{R}{\rho} \sin( \alpha)),
\end{equation}
where $\alpha$ is the angle between the interface and the y-axis as depicted in
Fig.~\ref{f:setup}. 
The negative sign is due to the downward direction of the
force. We remind the reader that $D$ is the depth of the quasi-two-dimensional setting so
that $F/D$ is the force per unit length. In the remainder we will therefore
use $D \gamma_b$ to make the force dimensionless. The radius of curvature for
the liquid interface, $\rho$, is equal to
\begin{equation}\label{e:rho}
\rho= \frac{L/2-R \sin \alpha}{\sin (\alpha -\theta)}.
\end{equation}
The final step is to convert equation~\ref{e:force} into a force-distance
relationship. For this, one needs to calculate the distance between the
center of mass of the particle and the interface, $y_p$, in terms of $\alpha$:
\begin{equation}\label{e:yp}
y_p= \rho(1-\cos (\alpha -\theta))-R\cos \alpha.
\end{equation}

Using equations~\ref{e:force}-\ref{e:yp}, the force-distance relationship is
theoretically calculated for the rigid particle and is compared to the force
measured from the molecular dynamics simulation in Fig.~\ref{f:rigid}(e). The
excellent agreement between the continuum theory and MD simulations in
Fig.~\ref{f:rigid}(e) indicates that the discrete molecular dynamics
simulations produce meaningful results to describe the behavior of the
particle-interface interaction on the continuum level. 

{
Fig.~\ref{f:rigid}(e) shows that as the particle moves up, the force exerted on
it by the interface almost linearly increases until $y_p/R=2$. Then, it starts
to saturate until it reaches the final maximum of 2. This value is expected
since the angle between the interface and the y-axis, $\alpha$, goes to $\pi$
during the movement of the particle. Inserting $\alpha=\pi$ and $\theta=\pi/2$
in Eq.~\ref{e:force} results in the final force to be two times $\gamma_b D$ in negative y-direction.
}

{
By fitting a circular arc to the interface shape, $\rho$ and $\alpha$ can be
obtained from the MD simulations.  The surface tension of the liquid-liquid and
particle-liquid interfaces are measured in a separate setup with a flat
interface, and similar interaction energies using the Kirkwood-Buff formula for
a planar interface.  This allows to compute the two components of the total
force, i.e., $F_c$, and $F_p$.  By comparing the two components of the force in
Fig.~\ref{f:rigid}, it can be concluded that the pressure force has a much
smaller role in the total force due to the relatively large bath. Furthermore,
the pressure force shows a non-monotonic behavior with $y_p$: as the particle
moves away from the interface the curvature of the bath interface increases
which results in more negative capillary pressure below the particle. At the
same time, the contact area between the particle and the lower liquid phase
decreases ultimately leading to a $F_p$ which reaches a maximum and then
decreases again. 
}

\section{Soft particles}
We now turn to the desorption of soft particles. First, we study the desorption
of a cylindrical elastic particle in the two-dimensional setup. We use a 2D setup, because computationally,
it is one order of magnitude faster to simulate a 2D particle as compared to its 3D counterpart.
Besides, the length of the contact line is fixed in the
two-dimensional setup and this actually facilitates the physical
interpretation. Next, we present the desorption profile of a spherical soft
particle from a square fluid interface. Due to the high computational cost
of these simulations, we limit ourselves to a small set in order to generalize
the results from a two to the three-dimensional configuration. Finally, we
present results for microgel particles (in 2D), in which the role of the
transition from the swollen to the collapsed state on the binding between the
particle and the interface is investigated. 

\subsection{Two dimensional setup}\label{s:2D}
%
%
%
\begin{figure}
\centering
\includegraphics[height=6.3cm]{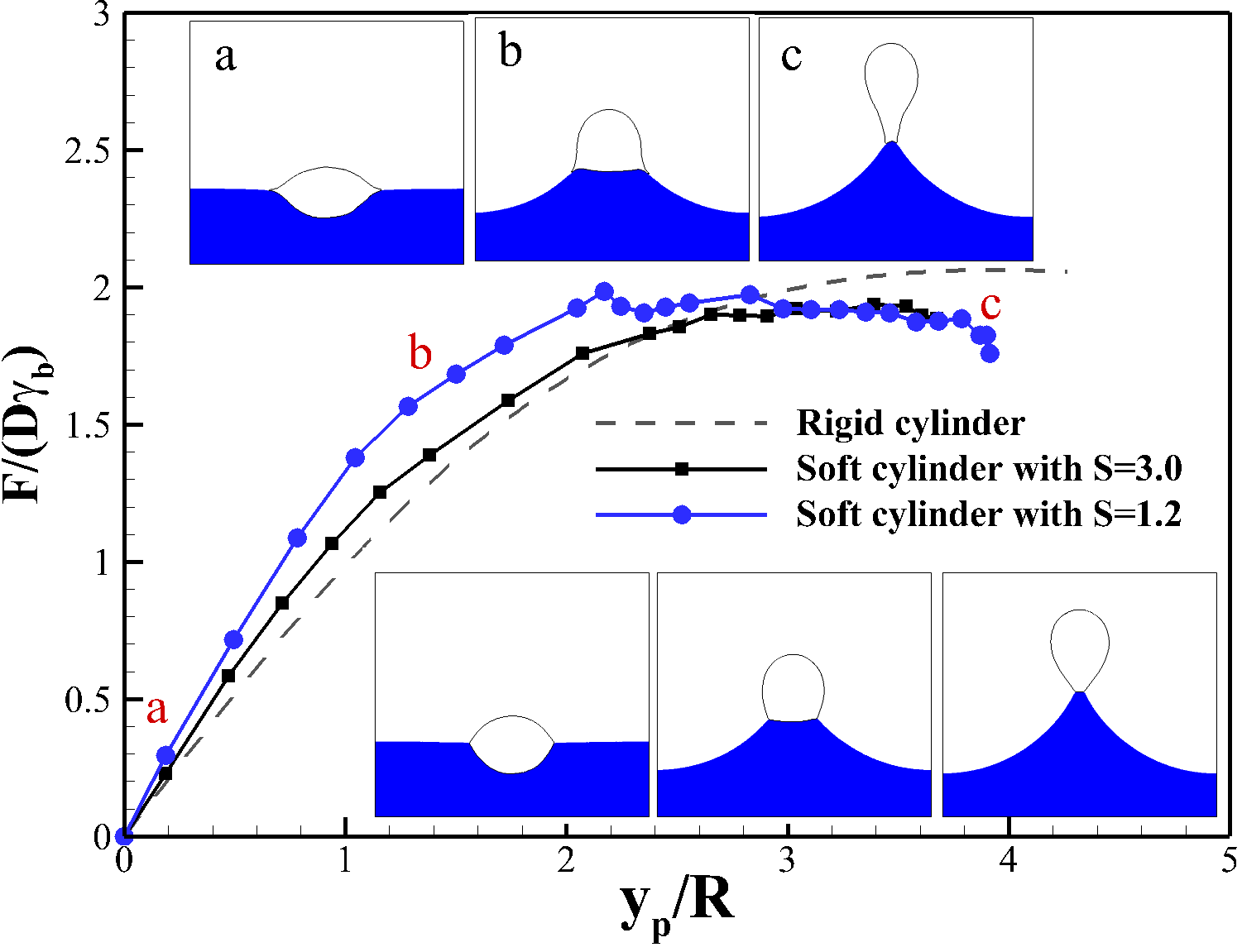}
\caption{The force-distance relationship for two soft cylindrical particles is
	compared to a rigid cylinder. Snapshots in the upper row show the
	desorption profile for the more deformable particle ($S=1.2$), while
	snapshots in the lower row depict the less deformable one ($S=3.0$).
	$S_b$ is kept at 3.1, and both particles are neutrally wetting.
	}
\label{f:2DS}
\end{figure}
%
%
%
Figure~\ref{f:2DS} shows the desorption profiles for two cylindrical elastic particles.  For
both cases there is no wetting contrast between the particle and the two liquid
phases ($\theta=90^\circ$) so both particles are initially top-down symmetric.
In both cases the bath elastocapillary number is the same, namely
$S_b=3.1$, so that the forcing by the bath compared to their elastic modulus is
identical. However, the particle elastocapillary numbers are
different, respectively $S=1.2$ and $S=3.0$. The latter has a larger
solid-liquid surface tension and thereby more strongly opposes deformation. The
leftmost snapshots of the insets in Fig.~\ref{f:2DS} show the equilibrium
positions of the particles, prior to applying the force. The upper row of
snapshots shows the more deformable particle ($S=1.2$) for which $\Delta R/R=
0.7$, while the lower row of snapshots shows the particle with a relatively
high solid-liquid surface tension ($S=3.0$), and hence, it has a smaller
deformation $\Delta R/R= 0.3$. After reaching the equilibrium shape at the
interface, particles are dragged up from the liquid interface quasi-statically,
by imposing a series of displacement and equilibration steps, and the
force-distance relationship is constructed. The result is shown in the main
panel of Fig.~\ref{f:2DS}, where each data point is obtained by averaging the
constraint force over 25 nanoseconds of equilibration run. 

As a reference case, the dashed line in Fig.~\ref{f:2DS} shows the
force-distance relationship for a rigid cylindrical particle with the same size
as the undeformed soft particle. Similar to rigid particles, the desorption
profile for the soft cylindrical particles can be roughly divided into two
stages. 
{At first, the restoring force increases almost linearly -- in this
phase the softest particle leads to the steepest slope of the
force-displacement curve. Hence, there is initially a larger restoring force
for softer particles. At the final stage of desorption, the restoring force
remains almost constant. A softer particle experiences a slightly weaker
maximum restoring force from the interface as compared to a rigid particle at this
stage. 
However, the soft particles detach at a shorter distance from the interface
than the rigid particle.
Interestingly, these two effects of softness, larger
force and earlier detachment, affect the desorption energy in the opposite
directions. Namely, the desorption energy is computed as the work done by the
force, and thus follows from the integration of the force-distance curve. When evaluated
quantitatively, the desorption energy is slightly smaller (less than 5\%) for the soft
particles.
}

%
%
%
\begin{figure}
\centering
\includegraphics[height=6.5cm]{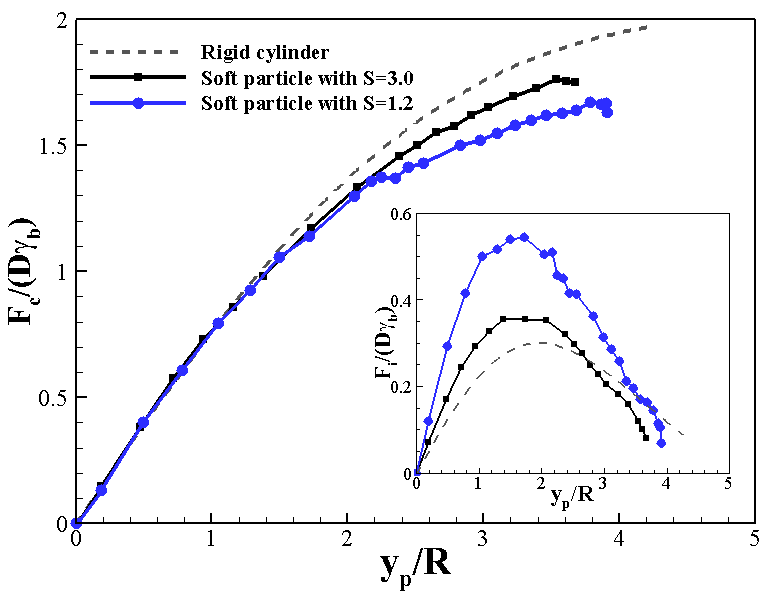}
\caption{The dimensionless form of the contact line force, $F_c$ and the
	pressure force $F_p$ (as the inset) are plotted for the soft
	cylindrical particles. Softer particles have a larger pressure force
	and smaller contact line force.
	}
\label{f:2DF}
\end{figure}
%
%

To understand the antagonistic roles of softness on the desorption energy of a
cylindrical particle better, it is insightful to plot the two constituent
forces acting on the particle, namely the pressure force $F_p$ and the contact
line force $F_c$. This is done in Fig.~\ref{f:2DF}, in which the contact line
force is plotted as the main curve and the pressure force is given as the
inset.  According to this figure, at the early stage of the desorption, the
contact line force is the same for both particles. However, it is smaller for
the soft particle at the later stage of the desorption. In contrast, the
pressure force, $F_p$, is larger for the softer particles. The combination of
these two behaviors results in the minor impact of softness on the total
desorption energy, as mentioned above.

{
These results can be rationalised as follows. In the early stage of the
desorption, the force exerted by the interface on the particle is mainly
parallel to the interface, i.e., it is horizontal, and thus deforms the particle
mainly in the x-direction. Therefore, the pressure force which is proportional
to the particle-interface contact area, i.e., $R+\Delta R $, is larger for the
soft particle. However, the force acting on the particle in the
y-direction is small, disabling it  to considerably deform the soft particle
normal to the interface. Therefore, the shape of the particle in the
y-direction is similar to the one of rigid particles.}

{
As the particle moves up, the interface bends upward, and the angle $\alpha$
increases, rendering the vertical component of the interfacial force larger.
Unlike the early stage, the normal restoring force is sufficiently large to
deform the particle now.  Therefore, rather than deforming the interface, it is
the soft particle that gets elongated in the y-direction. This hinders the
increase of the angle $\alpha$ and hence reduces the contact line force as
compared to the rigid particle. Recall that $\alpha$ is the angle between the
line tangent to the interface and the y-axis at the interface-particle contact
point. 
}

{
Changing the softness has no impact on the contact line force at the early
stage of the desorption. Instead, the increase in the overall restoring force
for soft particles in this regime is due to their larger contact area with the
interface. At the final stage of desorption, however, the opposite change in
the two components of the force balance each other resulting in an almost
constant force during the particle movement.
}

It would be of interest to generalise the observed force-distance relationship
to the case with a very large bath to particle size ratio: $L/R \gg 1$. Larger
$L/R$ results in a larger radius of curvature for the interface and hence,
smaller pressure force. Therefore, as the bath gets larger, the contribution of
the pressure force to the total force becomes negligible. Hence, the
force-distance relationship of the soft particle will essentially be due to
the contact line force. We showed that the deformation of the particle normal
to the interface decreases the contact line force. Although the extent of
reduction will be different for larger $L/R$, it is expected that an elastic
cylindrical particle always feels a smaller restoring force from the interface
as compared to the rigid particle. 

A further observation that needs to be discussed is the distance at which the
particle detaches from the interface. Unlike the rigid particle, a soft
particle gets elongated under the restoring force and thus, the distance
between its center of mass and the particle-interface contact line increases.
This can be easily seen by comparing the last snapshots for the two soft
particles in Fig.~\ref{f:2DS}. However, a soft particle is being reshaped by
the interface to keep the interface as flat as possible which requires the
particle-interface contact points to move faster toward each other, resulting
in early pinch-off of the particle. Therefore, the distance $y_p$ at which a
soft particle detaches from the interface depends on the outcome of its
horizontal and vertical deformation. In Fig.~\ref{f:2DS}, the more elastic
particle, the case with $S=1.2$, detaches at larger $y_p$ than the particle
with $S=3.0$, while its contact point with the interface is closer to the
interface. 

\subsection{Three dimensional setup}\label{s:3D}
%
%
%
\begin{figure}
\centering
\includegraphics[height=6cm]{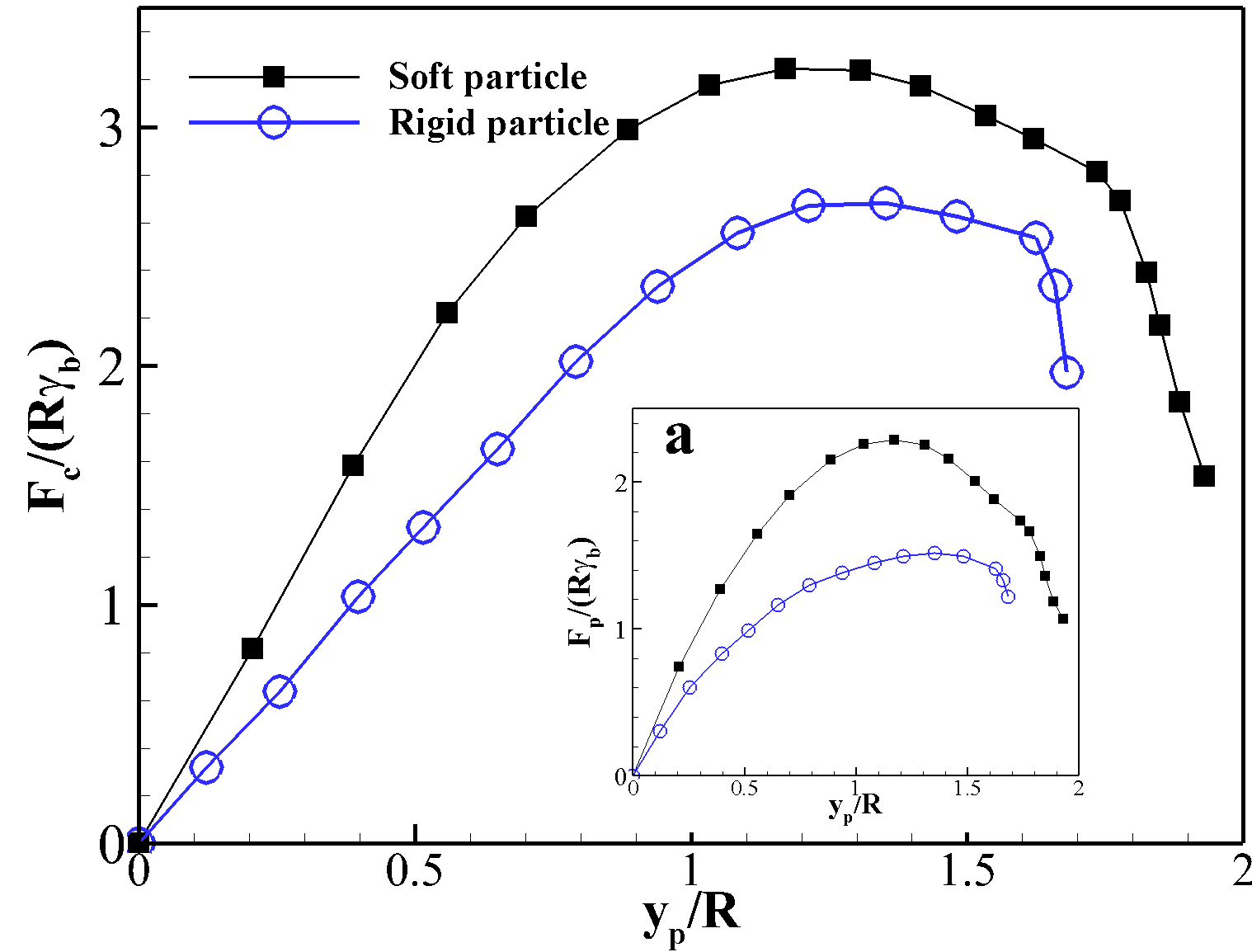}
\caption{The two components of the restoring force, $F_p$ and $F_c$, are compared
	for a rigid spherical particle and a soft particle with $\Delta
	R/R=0.4$. The contact line force increases for a soft spherical
	particle because its larger circumference compared to a rigid particle
	outweighs the decrease in the force during the detachment. The pressure
	force increases proportionally with the increase in the
	particle-interface contact area.}
\label{f:3DF}
\end{figure}
%
%
%
We now turn to the case of a soft spherical particle positioned at the center
of the square-shaped flat fluid bath ($D=L=6.4R$), as illustrated by the upper
row of snapshots in Fig.~\ref{f:logo}. The elastocapillary numbers for this
system are $S=0.9$, $S_b=1.5$, and the contact angle is $\theta=90^\circ$
(symmetric wetting). In simulations we first place the soft particle at the
interface and let it equilibrate for 30 nanoseconds, which results in the final
lenticular shape with the deformation of $\Delta R/R=0.4$. Then, we move the
particle's center of mass normal to the interface in a step-wise manner.  At
the same time we gather the restoring force incurred by the interface during
20-nanoseconds equilibration steps. Using the two-dimensional setup in
section~\ref{s:2D}, we showed that the pressure force increases as the softness
increases, almost proportional to the increase in the particle's wetted area,
while the contact line force decreases as the softness increases. Here, we
analyze how these features change from a two-dimensional configuration to the
three-dimensional one. 

The two components of the force are plotted in Fig.~\ref{f:3DF}. According to
this figure, unlike the 2D particle, the contact line force $F_c$, for
a soft spherical particle is larger compared to a rigid one. The reason for this
difference is that for the soft 2D particle, the length of the contact line remains
fixed during the deformation, while for a soft spherical particle, the length
of the contact line, which is proportional to $R+\Delta R$, increases when
compared to the rigid particle.  Therefore, the softness does affect the
contact line force in two opposite directions: the length of the contact line
and hence the resultant force increases, while the contact line force per unit
length decreases due to the deformability, similar to what happens to the soft
cylinder.  Hence, Fig.~\ref{f:3DF} reveals that the increase in the
circumference of the contact line is more prominent than the decreases in the
force per unit length. 

We now argue that the increase of contact line length
is always dominant, regardless of the particle to interface size ratio. For
this, we assume that the maximum deformation of the particle in the y-direction
is of the order of $\Delta R$, similar to its deformation at the interface.
Such an elongation decreases the vertical contact line force by reducing the
the angle $\alpha$, which determines the y-component of the force according to
Eqn.~\ref{e:force}. 
Using a simple geometrical analysis, it can be shown that the decrease in the
contact line force, $\Delta F_c/F_c$ is proportional to $\Delta R/L$ in which
$L$ is the liquid bath size. However, the increase in the particle
circumference, and hence the increase in the contact line force is proportional
to $\Delta R/R$. Since the liquid bath is generally larger than the particle,
we thus conclude that a soft spherical particle always experiences a stronger
contact line force as compared to the rigid particle. As such, in three
dimensions, the softness leads to a stronger attachment to the interface. 

\begin{figure}
\centering
\includegraphics[height=6cm]{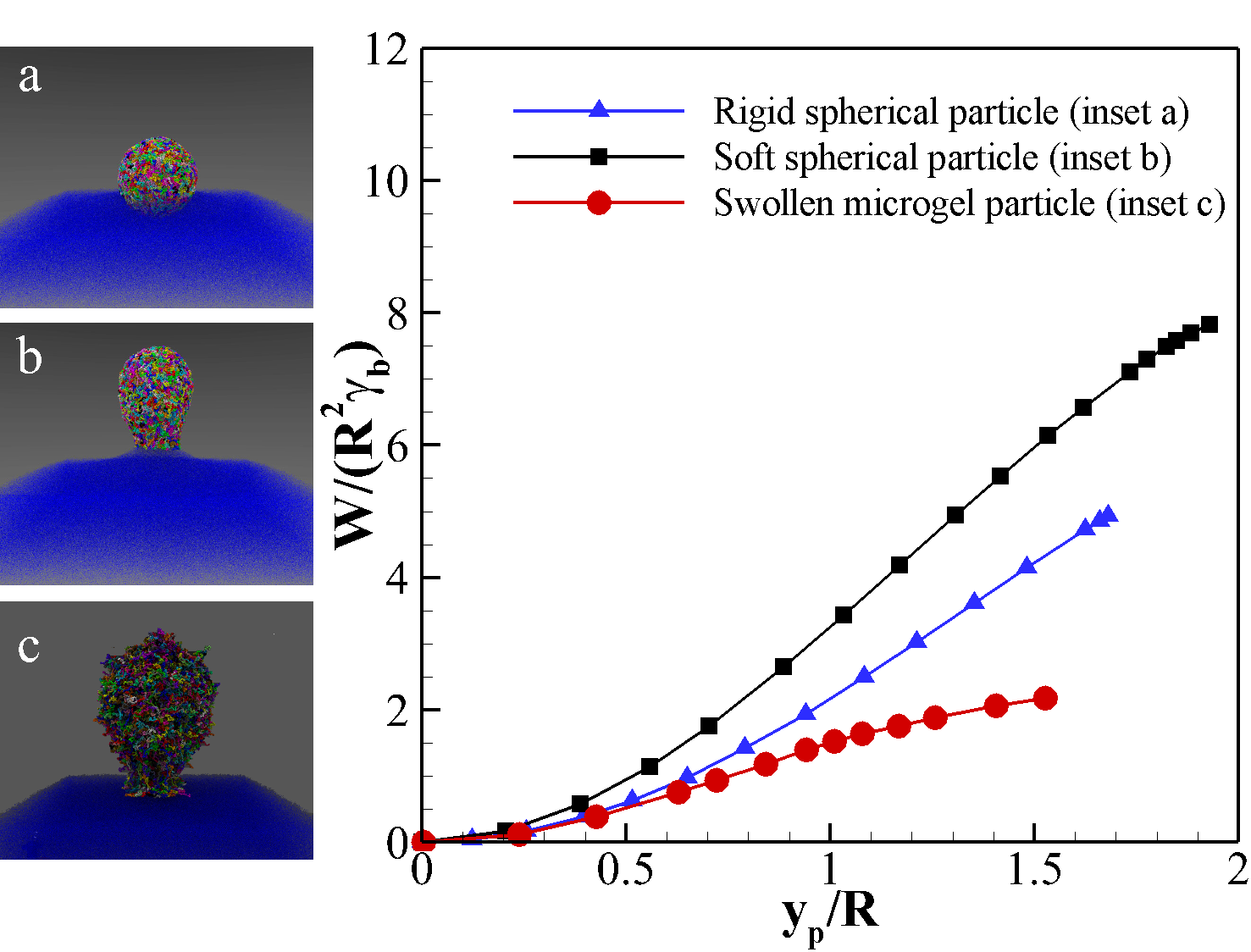}
\caption{Comparison of the work of detachment for a rigid spherical particle, a
	soft spherical particle (see the upper row of snapshots in
	Fig.~\ref{f:logo}), and a microgel particle (see the lower row in
	Fig.~\ref{f:logo}).}
\label{f:3DW}
\end{figure}

The pressure force $F_p$ is shown in the inset of Fig.~\ref{f:3DF}, for both
the soft and rigid spherical particle. The pressure force increases with the increase in
particle-interface wetting area. For a soft spherical particle, the
particle-interface contact area is approximately proportional to $(R+\Delta
R)^2$, while it was proportional to $R+ \Delta R$ for a deformed cylinder.
Therefore, it is expected that the soft particle experiences a larger increase
due to its deformation at the interface. For the soft spherical particle in
Fig.\ref{f:3DF}, the deformation is $\Delta R/R=0.4$, which results in an
increase of the pressure force by about a factor two as compared to the rigid
case.  Since both force components $F_c$ and $F_p$ increase for a soft
particle, the desorption energy turns out to be larger for soft particles
relative to a rigid particle with the same volume. This is shown in
Fig.~\ref{f:3DW}, in which the work of detachment from Eq.~\ref{e:E} is plotted
versus the particle's center of mass distance $y_p$ from the interface. 
%

\begin{figure}
\centering
\includegraphics[height=6cm]{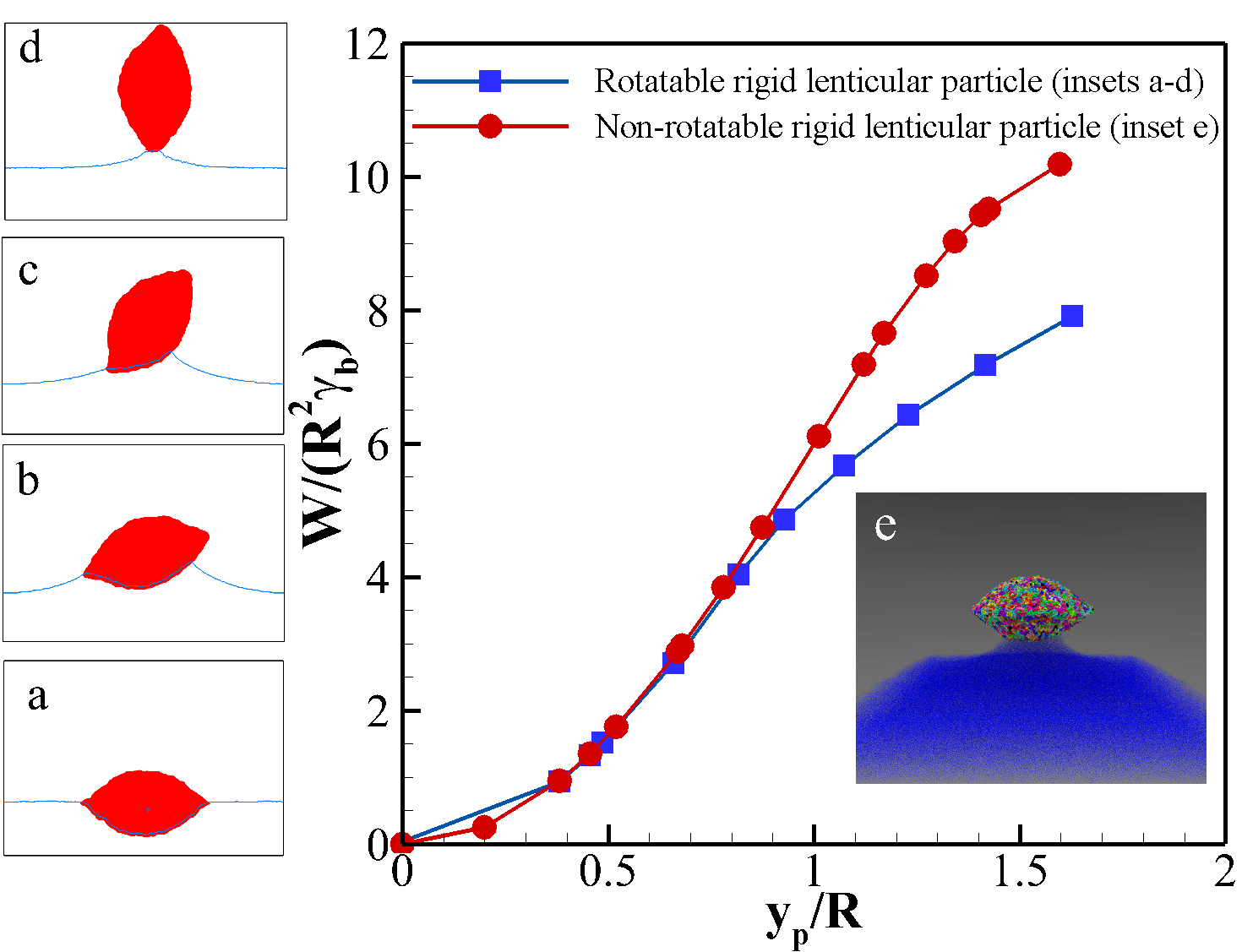}
	\caption{Comparison of the work of detachment for a rigid lenticular particle under two desorption scenarios, with and without particle rotation. Snapshots (a) to (d) illustrate the desorption when the particle can rotate. Snapshot (e) shows a particle with the rotation being disabled.
	}
\label{f:3DR}
\end{figure}

{
Although soft particles attach to the interface more strongly than their rigid counterparts, the antagonistic effect of softness reduces the expected increase in the desorption energy. Therefore, one could argue that a rigid lenticular particle with a shape similar to the deformed soft particle would result in higher desorption energy. To examine this point, we cross-link the deformed soft particle in Fig.~\ref{f:3DW}, converting it into a rigid lenticular particle with the same shape as the deformed soft particle at the interface. Then, the desorption free energy profile is reconstructed for this particle under two detachment scenarios. In the first scenario, we pull the particle's center of mass upward, similar to the other cases in this paper. The particle's behaviour during the desorption is shown in Fig.~\ref{f:3DR}, according to which the particle rotates during the desorption. This behaviour is in agreement with the observations for the desorption of non-spherical rigid particles \citep{lishchuk}. In the second scenario, we disable the rotational degree of freedom of the particle, i.e., it is pulled away from the interface while its main axis remains parallel to the interface. Desorption free energy profiles for these two cases are compared in Fig.~\ref{f:3DW}. According to this figure, for the non-rotated particle, the detachment work is much higher than for the soft particle, which agrees with our prediction. However, the detachment work for the rotated particle is very close to the soft particle.
}

Due to the computational limitation, the bath size is relatively small in our
simulations. It would be interesting to extend the results to cases with a very
large fluid bath. Then, the pressure force would be negligible as compared to
the contact line force and the contact line force determines the outcome. As we
discussed earlier, the contact line force is always larger for the soft
particle and therefore soft spherical particles have larger desorption energies
as compared to rigid particles.

If we use the same ratio $L/R=8.2$ of the two-dimensional to create a
three-dimensional setup, there are almost 25 times more beads in the system
corresponding to a more than one order of magnitude higher computational cost.
This is a demanding computation even for the largest available supercomputers.
Therefore, it is required to adopt the simulation parameters in order to reduce
the computational effort required for the simulations.  To achieve this we
chose a smaller box size of $L/R=D/R=6.4$, and the cut-off radius is reduced to
$r_c=2.5\sigma$ for three-dimensional simulations. With approximately 10
million beads in the simulated system, we perform about 600 nanoseconds of
simulated time for each three-dimensional particle.  On a modern GPU-based
supercomputing cluster based on Intel(R) Xeon(R) CPU E5-2450 v2 2.50GHz CPUs
and NVIDIA Tesla K40m GPUs, we are able to reach a performance of 1 nanosecond per hour. For this, our jobs run on 16 GPU nodes, where
each GPU is supported by 24 physical CPU cores.
\subsection{Microgel particles}\label{s:microgel}
Finally, we turn to microgel particles, which can exhibit different degrees of
swelling based on the chemical properties of the system. For example, a unique
feature of pNIPAM is that at a specific temperature, their structure transforms
from a swollen to collapsed state and vice versa. This means that at a certain
temperature polymer chains of the microgel particles becomes soluble in the
solvent, and increase their volume. It is critical to understand the change in
the particle properties between these two states. Motivated by the context of
emulsion stabilization, we now consider the desorption energy of microgel
particles for different degrees of swelling. 

On the molecular scale, the swollen to collapsed transition happens due to a
change in the intermolecular interactions between the solvent molecules and the
polymer chains. On the macroscopic level, this transition results in a few
differences. First, the size of the microgel particle increases in the swollen
state; a swollen microgel particle has a hydrodynamic diameter which is usually
2 to 3 times larger than its diameter in the collapsed state
\citep{Destribats2011,Tsuji2008}. Second, the swollen particle becomes more
deformable because the same amount of crosslinking spreads over a larger
volume. Third, the surface tension of the particle changes, and sometimes the
change is so drastic that an initially non-wetting liquid phase can become a
wetting phase \citep{Ngai2005}. 

\begin{figure}
\centering
\includegraphics[height=5.6cm]{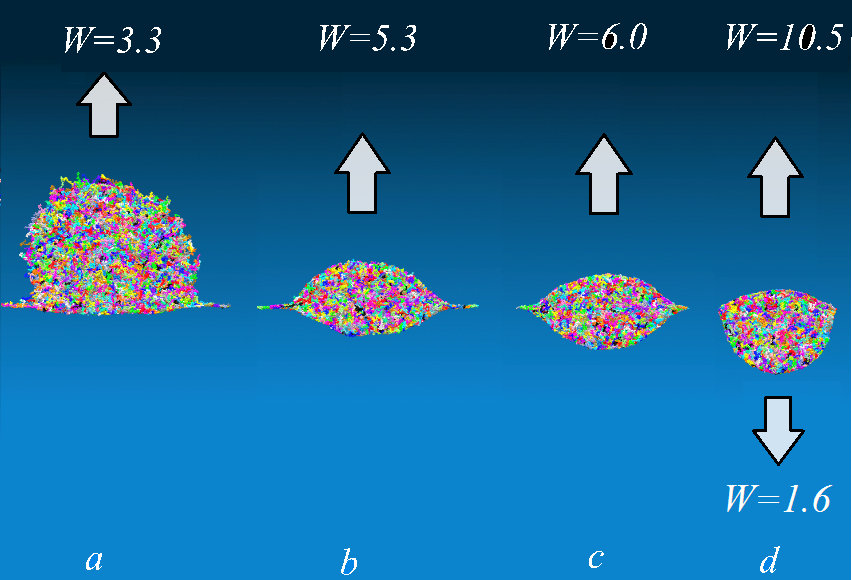}
\caption{Effect of the swollen to collapsed transition on the desorption energy
	of a two-dimensional microgel particle from a fluid interface. From
	left to right, the affinity between the microgel particle and the
	solvent molecules of the upper phase is increased, while all other
	parameters are kept fixed. The surface tension between the particle and
	the lower phase is $32.5 mN m^{-1}$, the surface tension of the bath is
	$83.5 mN m^{-1}$, and the density of the cross-linking for the particle
	immersed in the bulk of the nonwetting liquid is $0.8 nm^{-3}$. The
	radius of the microgel particle in the swollen state is about 1.9 times
	larger than its radius in the collapsed state. }
\label{f:swol-to-coll}
\end{figure}

In our molecular dynamics model, the physics of the problem is mimicked by
changing the Lennard-Jones interactions between the polymer chains of the
particle and the solvent molecules. For the soft particles studied in the previous sections, 
we increase the attraction between the
polymer chains and the upper liquid phase to the extent that polymer chains
become soluble in the upper liquid phase. As is shown in Fig.~\ref{f:swol-to-coll}(a), the resulting
microgel particle at equilibrium becomes asymmetric due to the
asymmetric interactions. 
In Fig.~\ref{f:swol-to-coll}, from left to right, we decrease the attractive Lennard-Jones term between the beads of
the polymer chains and the solvent molecules of the upper phase while keeping
the surface tension of the particle with the lower phase, the surface tension
of the bath, and the polymeric structure of the particle fixed. 
The corresponding $\beta$ parameters of Eqn.~\ref{e:LJ} which are used in these
simulations are presented in table~\ref{t:betas}.
The surface tension between the particle and the lower phase is
32.5 $mN m^{-1}$, the surface tension of the bath is 83.5 $mN m^{-1}$, and the
density of the cross-linking for the particle immersed in the bulk of the
nonwetting liquid is 0.8 $nm^{-3}$. The radius of the microgel particle in the
swollen state is 1.9 times larger than its radius in the collapsed state. 
 
\begin{figure}
\centering
\includegraphics[height=5.8cm]{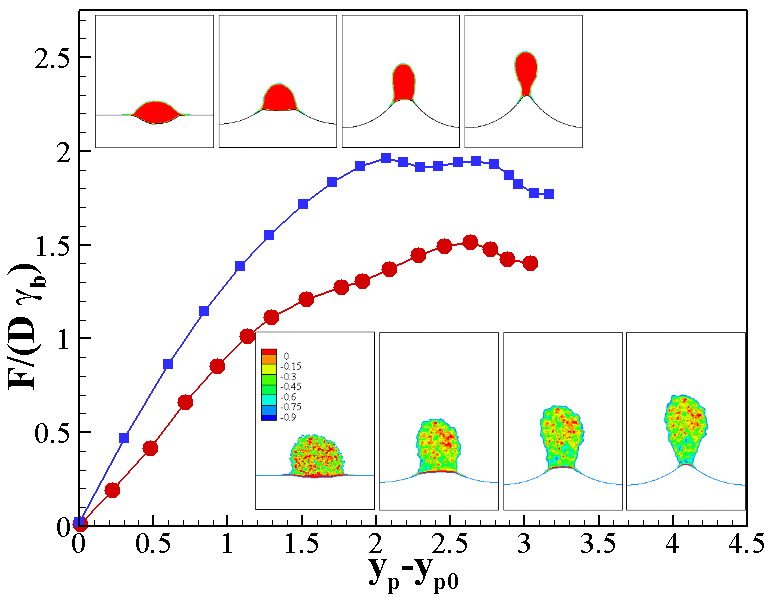}
\caption{The force-distance relationship is compared between a swollen microgel
	particle (lower inset) and a collapsed microgel particle (upper inset).
	The swollen microgel particle is the one shown in Fig.~\ref{f:swol-to-coll}(a)
	and the collapsed one is shown in Fig.~\ref{f:swol-to-coll}(b).
	The higher softness of the swollen microgel and its stronger affinity
	to the upper fluid phase weakens its binding to the interface.}
\label{f:microgel-2cases}
\end{figure}

Fig.~\ref{f:swol-to-coll}(a), shows a swollen microgel particle adsorbed to the
fluid interface. Due to the strong attraction between its polymer chains and
the solvent molecules of the upper phase, the polymer chains prefer to be
covered by the solvent molecules of the upper phase, resulting in a swollen
microgel particle. Also, it is energetically favorable to replace the
liquid-liquid contact at the bath interface with a particle-liquid contact.
Therefore, the swollen microgel particle gets adsorbed to the interface,
producing the typical fried-egg like shape (core-corona shape) of microgel
particles at a fluid interface\citep{Deshmukh2014a,Destribats2014,Monteillet2014}.
Fig.~\ref{f:swol-to-coll}(b) shows the same particle with a weaker
solvent-polymer affinity. The surface tension of the particle with the upper
phase is 7 $mN m^{-1}$. Here, there is a clear interface between the microgel
particle and the solvent. The particle is not swollen, but its size at the
interface is similar to the swollen case. Furthermore, the deformability of the particle
and the pulling force of the interface creates an
elongated particle whose size is comparable to the swollen case.
It is interesting to compare the desorption energy of this particle to the
one shown in Fig.~\ref{f:swol-to-coll}(a) to reveal the role of the chain
solvation on the particle-interface binding. Interestingly, the desorption free
energy is almost two times larger for the particle in
Fig.~\ref{f:swol-to-coll}(b). This suggests that the solvation of the polymer
chains of the microgel particle does not enhance its binding to the interface.
The reason for this behavior can be better understood by comparing the
desorption profile for two particles in Fig.~\ref{f:microgel-2cases}. As the
particle moves up, the swollen microgel deforms easily as compared to the
collapsed one, the interface remains horizontal, and hence the contact line
force in the y-direction becomes smaller for the swollen microgel particle. 
This is even more interesting if we consider the fact that the sizes of the
particles in Fig.~\ref{f:swol-to-coll}(a) and Fig.~\ref{f:swol-to-coll}(b) are
similar which means that the results can be easily generalized to
three-dimensional particles. In fact, the force-distance relationship for the
spherical swollen microgel particle shown in the second row of snapshots in
Fig.~\ref{f:logo} is presented in Fig.~\ref{f:3DW}. According to this figure,
the detachment work for the microgel particle is $W/ (R^2 \gamma_b)=2.2$. This is
much less than the detachment work of the rigid particle whose size is two
times smaller than the swollen microgel. Of course, this comparison is not
perfect because it is not possible to produce a rigid particle with the same
surface properties as the swollen microgel particle. However, it is interesting
to compare the desorption energy of a microgel particle with its polymer
chains adsorbed to the interface to a rigid particle with the same mass.

\begin{table}[]
\begin{tabular}{|c|c|}
	\hline
             & $\beta$        \\
	\hline
 Between lower and upper liquid phases            &	     0.0 \\
 Between lower phase and the particle     &	 0.80936     \\
 Between upper phase and the particle of inset a  &   0.94426\\
 Between upper phase and the particle of inset b  &   0.89030\\
 Between upper phase and the particle of inset c  &   0.80936\\
 Between upper phase and the particle of inset d  &   0.48562\\
 \hline
\end{tabular}
 \caption{Parameter $\beta$ of the Eqn.~\ref{e:LJ} between two liquid phases, and particle and liquid phases in Fig.~ \ref{f:swol-to-coll}.}
\label{t:betas}
\end{table}


%
%
%
%
%
%

Fig.~\ref{f:swol-to-coll}(c), shows a particle with a symmetric wetting. This case is important because a further decrease in the
affinity between the polymer chains and the solvent molecules of the upper
phase renders the desorption to the lower phase more favorable. The deformation
for this particle is $\Delta R/R=0.7$. Despite being less elongated than the
swollen microgel in Fig.~\ref{f:swol-to-coll}(a), its work of detachment is
almost two times higher suggesting that the collapsed microgel attaches to the interface more strongly than the swollen microgel particle. 

{
We can further decrease the attraction between the solvent molecules of the upper phase and the polymer chains of the particle to make the upper phase the non-wetting phase. Such a case is shown in Fig.~\ref{f:swol-to-coll}(d), the surface tension between the particle and the upper phase is increased to $\gamma=82.1 mN m^{-1}$ which results in $\Delta R/R=0.15$. We know that it is energetically favorable for the particle to move into the phase with which it has lower surface tension. Therefore, for the case in Fig.~\ref{f:swol-to-coll}(d), the particle moves into the lower phase, which is the wetting phase for this particle. The measured work of detachment verifies this point; it is 10.5 for the desorption into the upper phase, as compared to 1.6 for the lower phase.
In conclusion, the desorption energy for the swollen microgel is smaller than all other cases except for the case in which the direction of asymmetry gets reversed. This observation is consistent with experimental observations \citep{Ngai2005,Ngai2006}. Therefore, the difference between the binding free energy of the microgel particle in the swollen and collapsed state depends on the change in the surface properties.
}

{
Experimentally, it is observed that microgels are better stabilizers for emulsions \citep{Destribats2011}. One can think of two reasons for it. First, there is a stronger binding between a microgel particle and the fluid interface. Second, the particle-particle interactions at the interface get enhanced due to the entangled polymeric network of microgels \citep{Camerin2020a}. If we compare the free energy of desorption for a rigid particle in Fig.~\ref{f:2DF}, to those in Fig.~\ref{f:microgel-2cases}, we can conclude that the free energy of desorption for microgel particles is lower than for a rigid particle with the same mass. The same trend is observed for the spherical particles examined in Fig.~\ref{f:3DW}. Therefore, one would expect that the more enhanced stabilizing effect of the microgel particle comes from the particle-particle chain interactions.
}
%
%
%
%
%

{An important case which is not included in Fig.~\ref{f:swol-to-coll} is a microgel particle which is soluble in both phases, such as poly(N-vynilcaprolactam) microgel at the water-toluene interface. From computations, we found out that when such a particle moves away from the interface, it takes a long time for the interface to settle to its final shape; as a matter of fact, we were not able to obtain a reliable force measurement during reasonable simulation time. Also, the large size of such microgel particles requires a considerably larger computational box compared to other studied particles. These two reasons make their simulation computationally challenging. Therefore, the desorption of microgel particles that are soluble in both phases is a topic that needs to be investigated in a future study.}

\section{Conclusions}

Motivated by the need to understand how efficient soft particles are as
emulsion stabilizers, we measured the binding free energy between a soft
particle and an interface and compared it to the binding free energy of a rigid
particle with similar surface properties. In this study, the desorption of
cylindrical (two-dimensional) and spherical (three-dimensional) soft particles
is analyzed, and generic deformable particles, as well as soft polymeric
networks (microgel particles), are considered. 
We used coarse-grained molecular dynamics simulations with the thermodynamic
integration method to obtain the force-distance relationship for the particle
detachment.  We then measured the energy required to transfer a soft particle
from a fluid interface to the fluid bulk, the so-called desorption free energy. 

Our molecular dynamics simulations reproduce the analytical equation
for the force-distance relationship for the desorption of a rigid cylindrical
particle with $R=14\ nm $, proving the capability of the discrete nanoscale
simulations to capture the macroscopic continuum behaviour. 

We argued that to analyze the force-distance relationship, it is helpful to
split the total force acting on the particle during the desorption into two
major components: the contact line force, $F_c$, which pulls the particle
toward the interface along the particle-interface contact line and the pressure
force, $F_p$, which is produced by the curvature of the bath interface during
the desorption.

For a soft \emph{cylindrical} particle, the pressure force increases
proportional to the increase in the particle-interface contact area when
compared to a rigid particle. However, the contact line force decreases due to
the capability of the soft particle to deform instead of bending the interface.
Eventually, the desorption energy for a soft cylindrical particle is slightly
smaller than for a rigid particle with the same contact angle.

For a soft \emph{spherical} particle, the increase in the wetted area due to
particle deformability is more significant than the for soft cylindrical
particles. Therefore, soft spherical particles experience a larger increase in
the pressure force compared to the soft cylindrical particles.  Unlike soft
cylinders, soft spherical particles have a longer contact line compared to
rigid particles due to their three-dimensional circumference. Therefore,
despite having a smaller force per unit length as soft cylinders, the overall
contact line force increases for soft particles, and hence, a soft spherical
particle attaches to the interface stronger than a rigid particle.

For microgel particles, the affinity between the polymer chains of a swollen
particle and the solvent molecules of one liquid phase is decreased to
transform the swollen particle into a collapsed one. Using a simple area-based
argument, a collapsed microgel particle should be easier to detach due to its
smaller wetting area at the interface. However, the binding between the
particle and the interface can be stronger for a collapsed microgel due to its
larger surface tension and higher rigidity; the collapsed particle is less
deformable, and hence capable of bending the interface more effectively during
the desorption while the swollen microgel easily deforms, keeping the interface
nearly flat which creates a smaller pulling force. Also, the surface tension of
the particle and the liquid is larger for the collapsed microgel rendering it
less favorable to detach into the wetting phase. Finally, we showed that
depending on the the extent of change in the surface properties during the
swollen to collapsed transition, the particle can desorb to the initially
nonwetting phase, which could have a much smaller energy barrier than the newly
formed collapsed particle. 

In summary, the major conclusion from this work is that soft particles do not
show pronounced advantages to stabilize fluid-fluid interfaces when compared to
rigid particles. When taking into account the interplay of particle and
interface deformation, the binding free energy between an interface and a
particle would be less than a rigid particle with a similar shape as the deformed soft particle. 
In other words, our work suggests that an ellipsoidal particle could result in a stronger binding than a soft particle 
{if the particle rotation could be prevented,}
because the major axis of a rigid ellipsoid aligns with
the fluid interface, resulting in a higher reduction of interfacial area and
as such a stronger binding. Also, their shape will not change
during desorption which increases their capability to deform liquid interface
which is energetically less favourable. In addition, the  contact line pinning
could happen on such particles due to the slope variation on them.

\section{Acknowledgements}
We acknowledge access to the supercomputers Cartesius at the SURFsara
High-Performance Computing Center and Jureca at the J\"ulich Supercomputing
Centre. We acknowledge financial support from NWO through VIDI Grant No. 11304.

\bibliographystyle{rsc}

\providecommand*{\mcitethebibliography}{\thebibliography}
\csname @ifundefined\endcsname{endmcitethebibliography}
{\let\endmcitethebibliography\endthebibliography}{}

\end{document}